\documentclass{nature}

\usepackage{lineno}
\usepackage{amsmath,amssymb,graphicx,fancyhdr}
\usepackage{url}
\usepackage{caption}

\usepackage{graphicx}
\makeatletter
\let\saved@includegraphics\includegraphics
\AtBeginDocument{\let\includegraphics\saved@includegraphics}
\renewenvironment*{figure}{\@float{figure}}{\end@float}
\makeatother

\title{DISTRIBUTION AND HABITABILITY OF (META)STABLE BRINES ON PRESENT-DAY MARS.}

\author{Edgard G. Rivera-Valent\'{i}n$^{1,*}$, Vincent F. Chevrier$^{2,*}$, Alejandro Soto$^3$, \& Germ\'{a}n Mart\'{i}nez$^{1,4}$}

\begin{document}

\maketitle

\begin{affiliations}
 \item Lunar and Planetary Institute, Universities Space Research Association, Houston, TX 77058, USA.
 \item Arkansas Center for Space and Planetary Sciences, University of Arkansas, Fayetteville, AR 72701, USA.
 \item Southwest Research Institute, Boulder, CO 80302, USA.
 \item Department of Climate and Space Sciences and Engineering, University of Michigan, Ann Arbor, MI 48109, USA.
 \item[] $^{*}$ These authors contributed equally to this work
\end{affiliations}

\begin{abstract}
Special Regions on Mars are defined as environments able to host liquid water that meets certain temperature and water activity requirements that allow known terrestrial organisms to replicate\cite{Kminek:2010, Rummel:2014}, and therefore could be habitable. Such regions would be a concern for planetary protection policies owing to the potential for forward contamination (biological contamination from Earth). Pure liquid water is unstable on the Martian surface\cite{Ingersoll:1970, Haberle:2001}, but brines may be present\cite{Ingersoll:1970, Brass:1980}. Experimental work has shown that brines persist beyond their predicted stability region, leading to metastable liquids\cite{Gough:2011, Fischer:2016, Primm:2017}. Here we show that (meta)stable brines can form and persist from the equator to high latitudes on the surface of Mars for a few percent of the year for up to six consecutive hours, a broader range than previously thought\cite{Davila:2010, Martinez:2017}. However, only the lowest eutectic solutions can form, leading to brines with temperatures of less than 225~K. Our results indicate that (meta)stable brines on the Martian surface and shallow subsurface (a few centimeters deep) are not habitable because their water activities and temperatures fall outside the known tolerances for terrestrial life. Furthermore, (meta)stable brines do not meet the Special Regions requirements, reducing the risk for forward contamination and easing threats related to the exploration of the Martian surface. 
\end{abstract}

Here, using an experimentally validated thermodynamic framework for the (meta)stability of brines\cite{Chevrier:2009, Nuding:2014, Primm:2017, Primm:2018} paired with results from the Mars Weather Research and Forecasting (MarsWRF) general circulation model (GCM)\cite{Richardson:2007, Lee:2018}, we investigated when and where brines may form and persist for at least an hour on the surface of present-day Mars. Specifically, brine formation through deliquescence, the transition from a solid crystalline salt into an aqueous solution, as well as the stability of melt-induced brines against freezing, evaporation, and boiling were studied. If these (meta)stable liquids surpassed temperature and water activity limits of $T>250$~K and $a_{w}>0.5$, respectively\cite{Kminek:2010}, or stricter limits of $T > 255$~K and $a_{w} > 0.6$, as set by a later study\cite{Rummel:2014}, their host environments were considered Special Regions. Because these limits are set by the known limits of life in extreme environments on Earth, such regions on Mars could be habitable to some terrestrial organisms. 

First, we considered the stability of a brine regardless of formation mechanism or composition. To be stable against evaporation, the difference between the atmospheric water vapor ($e$) and the water vapor pressure just above the brine ($e_{b}$) must be minimized, such that $e_{b}\approx e$. Because water activity is defined as $a_{w}=e_{b}/p_{sat,l}$, where $p_{sat,l}$ is the saturation vapor pressure above pure liquid water, then when the brine is stable against evaporation $a_{w} = e/p_{sat,l}$, which is simply relative humidity with respect to liquid. In Figure 1, the resulting water activities of stable solutions over the possible Martian surface $T$ and $e$ are shown. Above $\sim270$~K, brines are not stable against evaporation. At these temperatures, the water activity required such that $e_{b}\approx e$ is below the lowest water activity observed for even a metastable single component brine\cite{Nuding:2014}. For $T < 230$~K, saturation with respect to water ice occurs (i.e., $RH_{i}\geq100\%$); depending on the eutectic temperature ($T_{E}$) of the brine, ice and a metastable liquid is possible. As an example, the metastable region for calcium perchlorate brines, which have the lowest known eutectic temperature ($T_{E}\approx 198$~K)\cite{Pestova:2005} for single component solutions of Mars-relevant salts, is shown in Figure 1. At the intermediate temperatures (i.e., $T_{E} < T < 270$~K), (meta)stable brines are possible. Indeed, at these temperatures $p_{sat,l}$ is smaller than the typical low atmospheric pressures on Mars (e.g., $p_{sat,l}$(230~K) = 14~Pa and $p_{sat,l}$(270~K) = 484~Pa) and so boiling is not favored. This regime, where brines are stable against boiling, evaporation, and freezing, constrains the salts that can lead to stable liquids on Mars. Based on the possible $T$ and $e$ pairings, the highest water activity for a stable brine is 0.66; therefore, brines with a higher eutectic water activity are not stable on present-day Mars (see Supplementary Figure 2 for further details). 

Of the Mars-relevant hygroscopic salts, magnesium perchlorate (Mg(ClO$_4$)$_{2}$) and calcium perchlorate (Ca(ClO$_4$)$_2$) both have eutectic $a_{w}\leq0.66$ (ref. \cite{Chevrier:2012}). Here we considered brine formation through deliquescence of magnesium perchlorate with a eutectic temperature and water activity of $T_{E} = 206$~K and $a_{w} = 0.56$ (ref. \cite{Chevrier:2009}) respectively, and calcium perchlorate with eutectic values of $T_{E}=198$~K and $a_{w}=0.52$ (ref. \cite{Pestova:2005}), across the surface of Mars. In order for a salt to deliquesce, the temperature must be above the eutectic and the ambient relative humidity with respect to liquid ($RH_{l}$) must be above the temperature-dependent deliquescence relative humidity ($DRH$). Though thermodynamically solutions should recrystallize once $RH_{l}<DRH$, due to a hysteresis effect, efflorescence occurs at a lower relative humidity\cite{Gough:2011, Nuding:2014}, allowing for metastable liquids. For example, though calcium perchlorate requires high relative humidity to deliquesce, once in solution it can effloresce, depending on temperature, at $RH_{l}\approx1\%$ (ref. \cite{Nuding:2014}). Furthermore, experiments under Mars-like conditions have shown that solutions persist even after saturation with respect to water ice is reached\cite{Nuding:2014, Primm:2017}, up to $RH_{i}\approx145$\%. In this case, if the local temperature is above the eutectic, an ice and metastable liquid mixture is present. We show the phase diagram including both stable and metastable regions of calcium perchlorate in Supplementary Figure 3.

Using these experimental constraints on brine (meta)stability, in Figure 2 we mapped the distribution of (meta)stable brines on the surface of Mars. Magnesium perchlorate brines are restricted to the northern hemisphere above 50$^{\circ}$N where they may be (meta)stable for up to 0.2\% of the year for at most eight hours per sol. Typically, though, these brines only persist for some 2~hrs/sol for $\sim0.04$\% of the year (i.e., a total of $\sim$6~hrs per year). Overall, such brines would be (meta)stable over 5\% of the Martian surface around a solar longitude (Ls) of 160$^{\circ}$. On the other hand, Ca(ClO$_4$)$_2$ brines can form and persist over some 40\% of the Martian surface, including near the equator. These brines are most likely at high northern latitudes, where they are (meta)stable for up to 2\% of the year. Although Ca(ClO$_4$)$_2$ brines may persist for at most half a sol per sol, they may be liquid for at most six consecutive hours. Typically, these brines occur transiently ($\sim3$~hrs/sol) for some 0.4\% of the year during the summer. In the northern hemisphere, Ca(ClO$_4$)$_2$ brines form and persist around Ls 140$^{\circ}$ while in the southern hemisphere around Ls 225$^{\circ}$. 

To constrain the habitability of the most favorable case, in Figure 3a we mapped the maximum water activity attained by a calcium perchlorate brine formed through deliquescence with the corresponding brine temperature in Figure 3b. The maximum brine $a_{w}$ is reached while the solution is metastable with ice. The maximum water activity of such brines is $a_{w}\approx0.8$ with a corresponding temperature of $\sim$205~K. On the other hand, the maximum temperature in Figure 3b of 210~K is associated with a water activity of 0.77. In Supplementary Figure 4, we show the maximum brine temperature and associated water activity. These conditions are met just prior to efflorescence. Though a brine may reach 225~K the associated $a_{w}$ is only 0.24. 

Furthermore, using updated and recalibrated in-situ measurements by the Phoenix lander\cite{Fischer:2019} paired with a model of the subsurface environment\cite{RiveraValentin:2018}, we investigated brine formation in the presence of a shallow ice table 10 cm deep\cite{Piqueux:2019} and the resulting brine properties in the regolith above the ice. In Supplementary Figure 5, we plot the conditions during which calcium perchlorate would deliquesce and a resulting brine would persist. Such liquids would have a maximum brine temperature of 217~K with a corresponding $a_{w}=0.21$, while the maximum water activity is 0.80 with a corresponding temperature of $T=206$~K. Such brines would persist longer than at the surface, lasting 5\% of the martian year at 5~cm and up to some 10\% of the martian year at 8~cm. 

The kinetics of deliquescence may be too slow to form liquids under the low Martian temperatures\cite{Fischer:2014}. Melting through salt-ice interactions, though, has been shown to occur within minutes of reaching the eutectic temperature\cite{Fischer:2016}. Thus, melting may be the preferred liquescence process on present-day Mars, potentially by salts interacting with seasonal surface frost or with subsurface ice tables. The maximum water activity reached by brines formed through melt of surface frost is captured in Figure 3 because such brines form when $RH_{i}\geq100$\% and thus when surface conditions would be in the ice and liquid metastability region. For brines formed by salt contact with a shallow ice table, the main water vapor source is the ice table, such that $e = p_{sat,i}(T)$, where $p_{sat,i}$ is the saturation vapor pressure above water ice. These brines would then have at most a water activity of $a_{w}=p_{sat,i}/p_{sat,l}$; thus, for their water activity to surpass the Special Region requirement, $T>190$~K or $T>205$~K for a limit of $a_{w}=0.5$ and $a_{w}=0.55$ respectively\cite{Kminek:2010, Rummel:2014}. Shallow ice tables a few centimeters deep, though, predominantly exist poleward of 50$^{\circ}$ latitude on Mars\cite{Mellon:2004, Piqueux:2019}, where the average annual surface temperatures are $\sim$200~K. Additionally, the maximum diurnal temperature significantly decreases with depth, with typical diurnal skin depths of 5~cm (ref. \cite{Mellon:2004}). Thus, though brines formed in the subsurface through salt-induced melting of the ice table may form and persist diurnally\cite{Fischer:2016}, their habitability would be largely temperature limited. 

Indeed, at the Special Region limit a stable brine would require a local partial pressure of water vapor of $e = 47$~Pa ($T=250$~K, $a_{w}=0.5$)\cite{Kminek:2010} or $e=88$~Pa  ($T=255$~K, $a_{w}=0.6$)\cite{Rummel:2014}. This is much higher than is available at the Martian near-surface\cite{Mellon:2004, Smith:2004}, where in situ measurements at contradistinct latitudes have peaked at $e\sim1$~Pa (ref. \cite{RiveraValentin:2018, Fischer:2019}). Nevertheless, would such a brine form, it would evaporate at some 20~$\mu$m/hr and 50~$\mu$m/hr, respectively, and thus would persist for much less than an hour considering even the total Martian atmospheric column abundance of water vapor\cite{Mellon:2004, Smith:2004}. 

Our work here has focused on pure phases while most naturally occurring salt deposits on Mars are salt mixtures. In this case, depending on the composition, the resulting eutectic temperature of the brine will be affected. Indeed, multicomponent brines have systematically lower $T_{E}$ and $DRH$ than pure phases\cite{Gough:2014}. Although such brines could amplify both the global distribution and duration of liquids on Mars, they would not increase their habitability because such liquids would be forming under colder temperatures and lower water activities. Another consideration is local scale effects. The GCM employed here has a resolution of 5$^{\circ}\times$5$^{\circ}$ and so averages over smaller, local scale effects (e.g., topography and terrain differences). Again, such effects would primarily impact the spatial and temporal distribution of brines. 

Our results show that metastability expands the locations and duration of brines on Mars, beyond what was previously thought\cite{Davila:2010, Martinez:2017} by including some equatorial regions. Various observations have indicated that brines may be presently forming on Mars (e.g., recurring slope lineae\cite{Chevrier:2012, Dundas:2017, Wang:2019}, and in-situ environmental data\cite{Primm:2018, RiveraValentin:2018}), though the only direct observation of liquids has been on the Phoenix lander strut as droplets that formed under the spacecraft-induced warmed environment\cite{Renno:2009}. Considering the distribution of active mass movement events on Mars, such as recurring slope lineae, calcium perchlorate brines, or brines with similar properties, may act as a trigger mechanism\cite{Dundas:2017, Wang:2019} due to their predicted broader distribution compared to magnesium perchlorate brines. However, should future experiments demonstrate that the timescale for deliquescence under Mars-relevant conditions is much longer than six hours, then another trigger mechanism for these mass movement events would be required.  

Considering planetary protection, our results indicate though high water activity solutions ($a_{w}\geq0.6$) may be stable on present-day Mars, the corresponding brine temperature is systematically below 210~K, which is well below the Special Regions temperature requirement and temperature limits for life\cite{Clarke:2013}. While brines with temperatures up to 225~K may form, their $a_{w}<0.25$, which is well below the Special Regions water activity requirement and known water activity limits for life\cite{Stevenson:2017}. Our work shows that stable or even metastable brines do not simultaneously attain the conditions required for their locations to be considered a Special Region. This is because of Mars' hyperarid conditions, which require lower temperatures to reach high relative humidities and tolerable water activities. Indeed, the expected maximum brine temperature (225~K) is at the boundary of the theoretical low temperature limit of life\cite{Clarke:2013}; thus, present-day conditions on Mars result in brine habitability being temperature limited. Consequently, the expected properties of (meta)stable brines on the surface and shallow subsurface of Mars are not habitable to known terrestrial life, which reduces the potential for forward contamination of the Martian surface.  

\section*{References}

\bibliographystyle{naturemag}

\newpage

\begin{addendum}
\item[Corresponding Author] Edgard G. Rivera-Valent\'{i}n~(email: rivera-valentin@lpi.usra.edu).
 \item This material is based on work funded by NASA under Grant No. 80NSSC17K0742 issued through the Habitable Worlds program and partially under Grant No. NNX15AM42G issued through the Mars Data Analysis program. The authors thank Christopher Lee for help with the microphysics scheme in MarsWRF and Jennifer Hanley for valuable feedback. The Lunar and Planetary Institute (LPI) is operated by Universities Space Research Association (USRA) under a cooperative agreement with the Science Mission Directorate of NASA. LPI Contribution No. 2322.
 \item[Author Contributions] E. G. R.-V. wrote software to automate the pairing of the salt stability fields with the general circulation model outputs, analyzed the data, created the figures, and drafted and revised the manuscript. V. F. C. conceived the original concept for this Letter and developed the stability fields for calcium and magnesium perchlorate. A. S. set up and ran the MarsWRF model, acquired and reduced GCM results, and drafted the relevant methods section. G. M. provided the recalibrated environmental data from the Phoenix lander, reduced the data, and aided in its interpretation. All authors contributed to the interpretation of the results and revision of this manuscript. 
\item[Competing Interests] The authors declare that they have no competing financial interests.
\end{addendum}

\newpage

\begin{methods}
\subsection{General Circulation Model.}
For this study, we used the Mars Weather Research and Forecasting (MarsWRF) general circulation model, a Mars version of the National Center for Atmospheric Research's Weather Research and Forecasting model\cite{Skamarock:2008, Richardson:2007, Toigo:2012}, which has been vetted against measurements by Viking and the Thermal Emission Spectrometer (TES) on the Mars Global Surveyor\cite{Lee:2018}. MarsWRF solves the primitive equations on a finite difference mesh using an Arakawa C-grid. For surface temperature calculations, a multilayer subsurface thermal diffusion and surface energy balance model uses surface albedo and thermal inertia maps derived from orbital observations of the Martian surface by TES, with water ice albedo and emissivity set to 0.33 and 1.0, respectively. For radiation calculations, we use a two-stream radiation code that implements a k-distribution radiative transfer scheme (see \cite{Mischna:2012}). The total present-day atmospheric CO$_{2}$ budget has been tuned such that modeled pressure curves match the observed pressure curves at the Viking 1 and Viking 2 landing sites. MarsWRF simulates the sublimation, condensation, sedimentation, and transport of water ice particles in the atmosphere. This version uses the water cycle and radiatively-active microphysics scheme implemented by Lee et al. (2018)\cite{Lee:2018}.

For this work, simulations were run at 5$^{\circ}$ by 5$^{\circ}$ horizontal resolution with 52 vertical levels. The vertical grid stretches from $\sim$75-100~m above the ground, depending on location and season, up to 120~km; this grid uses a modified-sigma, terrain-following vertical coordinate. We ran the simulations for a total of six Martian years. For the final year, we output hourly surface conditions.

\subsection{Brine Modeling}
The hourly surface conditions, specifically surface temperature ($T_{s}$), surface pressure ($P$), and water vapor specific mixing ratio ($\gamma_{wv}$), from MarsWRF were used in the thermodynamic analysis. For every simulated location on the surface of Mars and for every modeled hour, the surface water vapor specific mixing ratio was converted to water vapor pressure by 
\begin{equation}
e = \frac{\gamma_{wv} P}{\mu + \gamma_{wv}}
\end{equation}
where $\mu$ is the ratio of the gas constants for dry and moist air. Surface relative humidity with respect to liquid ($RH_{l}$) is then found by $RH_{l} = 100\left[e/ p_{sat,l}\left(T_{s}\right)\right]$, where $p_{sat,l}$ is the temperature-dependent saturation vapor pressure above pure liquid water, here following the formulation from\cite{Murphy:2005}. Ice formation at the surface can occur when saturation with respect to ice is reached (i.e., $RH_{i}=100\%$). Relative humidity with respect to liquid is related to relative humidity with respect to ice as $RH_{i} = RH_{l} \left[p_{sat,l}\left(T_{s}\right)/p_{sat,i}\left(T_{s}\right)\right]$, where $p_{sat,i}$ is the saturation vapor pressure above ice, here following the formulation by\cite{Feistel:2007}. For every hour, the deliquescence relative humidity ($DRH$) was calculated for calcium perchlorate\cite{RiveraValentin:2018} and magnesium perchlorate\cite{Primm:2018}. If $RH_{l}\geq DRH$ and the surface temperature surpassed the brine's eutectic temperature, a brine was considered to have formed. The properties of the brine (e.g., temperature and water activity) along with the time in solution were tracked until either $RH_{i}>145$\% (ref. \cite{Primm:2017}), at which point the solution would entirely freeze out, or the efflorescence relative humidity was reached, at which point the salt recrystallizes and water is lost to the ambient atmosphere, thus accounting for metastable brines. 

The water activity ($a_{w}$) of a brine is by definition $a_{w}=e_{b}/p_{sat,l}$, where $e_{b}$ is the water vapor pressure directly above the brine. If $e_{b}>e$, the brine will lose water via evaporation, which on Mars would follow the Ingersoll formulation\cite{Ingersoll:1970}, until $e_{b}$ approaches $e$. When the brine is at equilibrium with the ambient air, $a_{w}=e/p_{sat,l}$ and thus $a_{w} = RH_{l}/100$. Such a relation has been extensively verified by Mars-relevant evaporation experiments (e.g., \cite{Chevrier:2008, Chevrier:2009, Hanley:2012}), and is commonly used for food storage and preservation regulation (e.g., \cite{Stevenson:2017} and references therein). Stability against boiling was checked ensuring $e_{b}\leq P$. 

\subsection{Subsurface Modeling}
Newly recalibrated environmental data from the Thermal and Electrical Conductivity Probe (TECP) on the Phoenix lander\cite{Fischer:2019} were used to validate our subsurface environmental modeling, following the methods described in \cite{Chevrier:2012, Kereszturi:2012, RiveraValentin:2018}. The subsurface temperature and humidity environment were simulated by solving the 1-D thermal and mass diffusion equations via finite element analysis with element size of 1 cm and time step of 26~s. We modeled to a depth of 4~m, which is several times the annual skin-depth, and simulated three Martian years, saving the results of the last year. Running over multiple years ensures the code converges to a stable solution. We assumed regolith properties of $\Gamma = 150$~J~m$^{-2}$~K$^{-1}$~s$^{-1/2}$, $A=0.18$, $\varphi=0.16$, and $\tau = 2$ for thermal inertia, albedo, porosity, and tortuosity respectively\cite{Hudson:2008, Sizemore:2008, Savijarvi:2020}. Environmental conditions per element were saved at hourly increments. Simulated surface conditions were validated against Phoenix TECP temperature and relative humidity measurements. 
\end{methods}

\begin{addendum}
\item[Data availability] The data that support the figures within this paper are available at \url{https://doi.org/10.6084/m9.figshare.11907984} or from the corresponding author upon reasonable request. The newly recalibrated environmental data from the Thermal and Electrical Conductivity Probe on the Phoenix lander are available on the NASA Planetary Data System Geosciences Node (\url{https://pds-geosciences.wustl.edu/missions/phoenix/martinez.htm}).
\item[Code availability] The MarsWRF (Mars Weather Research and Forecasting) general circulation model is available from A. S. upon reasonable request.
\end{addendum}

\begin{figure}
\includegraphics[width=\textwidth]{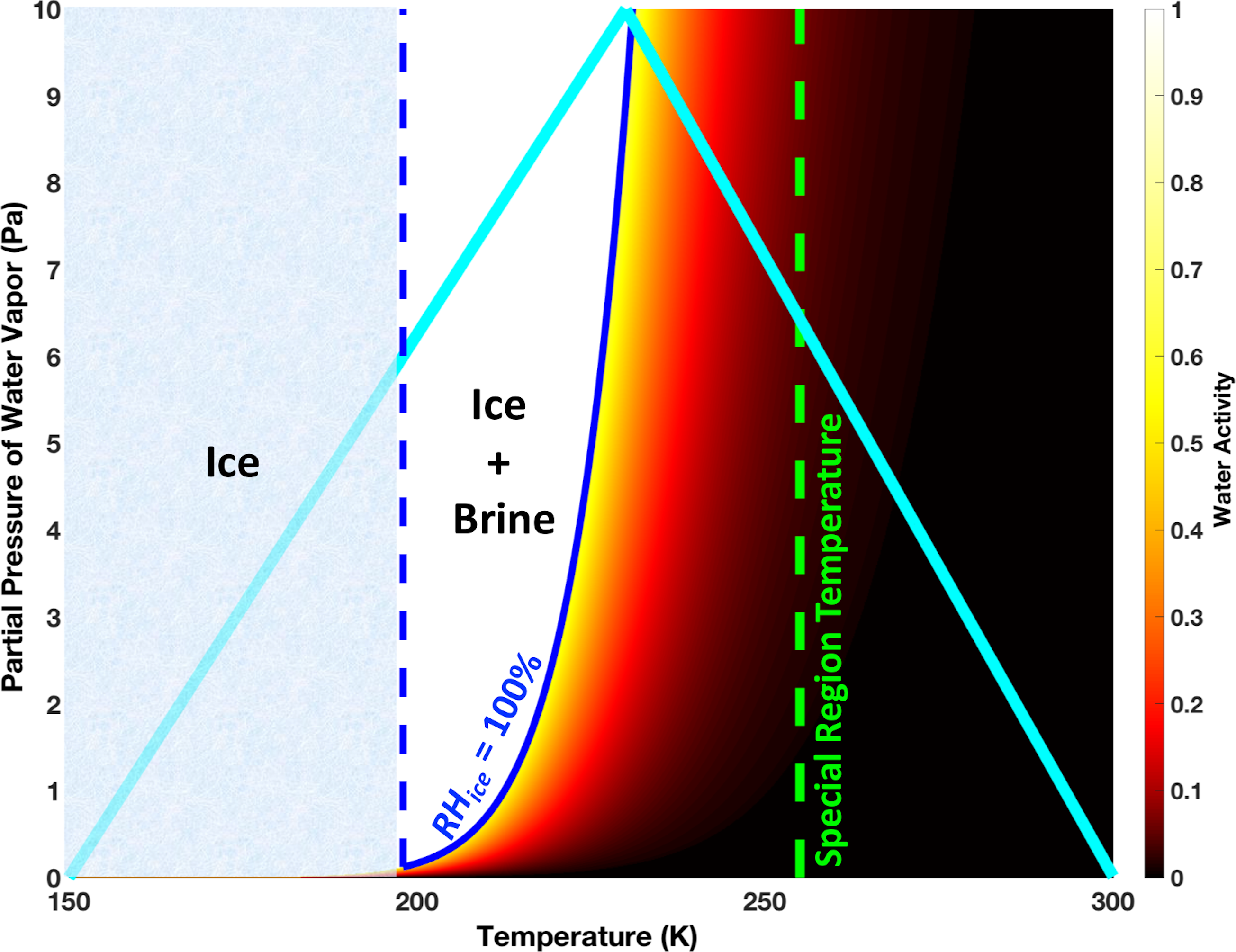}
\caption{\small{\textbf{Water activity of stable brines over Mars-relevant temperatures and water vapor pressures.} Over the range of possible global surface temperature and water vapor pressure on Mars, the resulting water activity of an arbitrary stable solution is shown here in color. The Mars-relevant pairings of surface temperature and water vapor are contained between the cyan lines (see Supplementary Figure 1 for details). The solid blue line is where saturation with respect to ice is achieved (i.e., where relative humidity with respect to ice is $RH_{ice}=100$\%). However, depending on the eutectic temperature of the brine, a metastable liquid and ice solution may be present. Here we show the eutectic temperature of a calcium perchlorate brine (dashed blue line) as an example. Only ice will exist below the eutectic temperature of the brine. Additionally, the Special Region lower limit temperature\cite{Rummel:2014} ($T = 255$~K) is noted by the green dashed line.}}
\end{figure}

\newpage

\begin{figure}
\includegraphics[width=\textwidth]{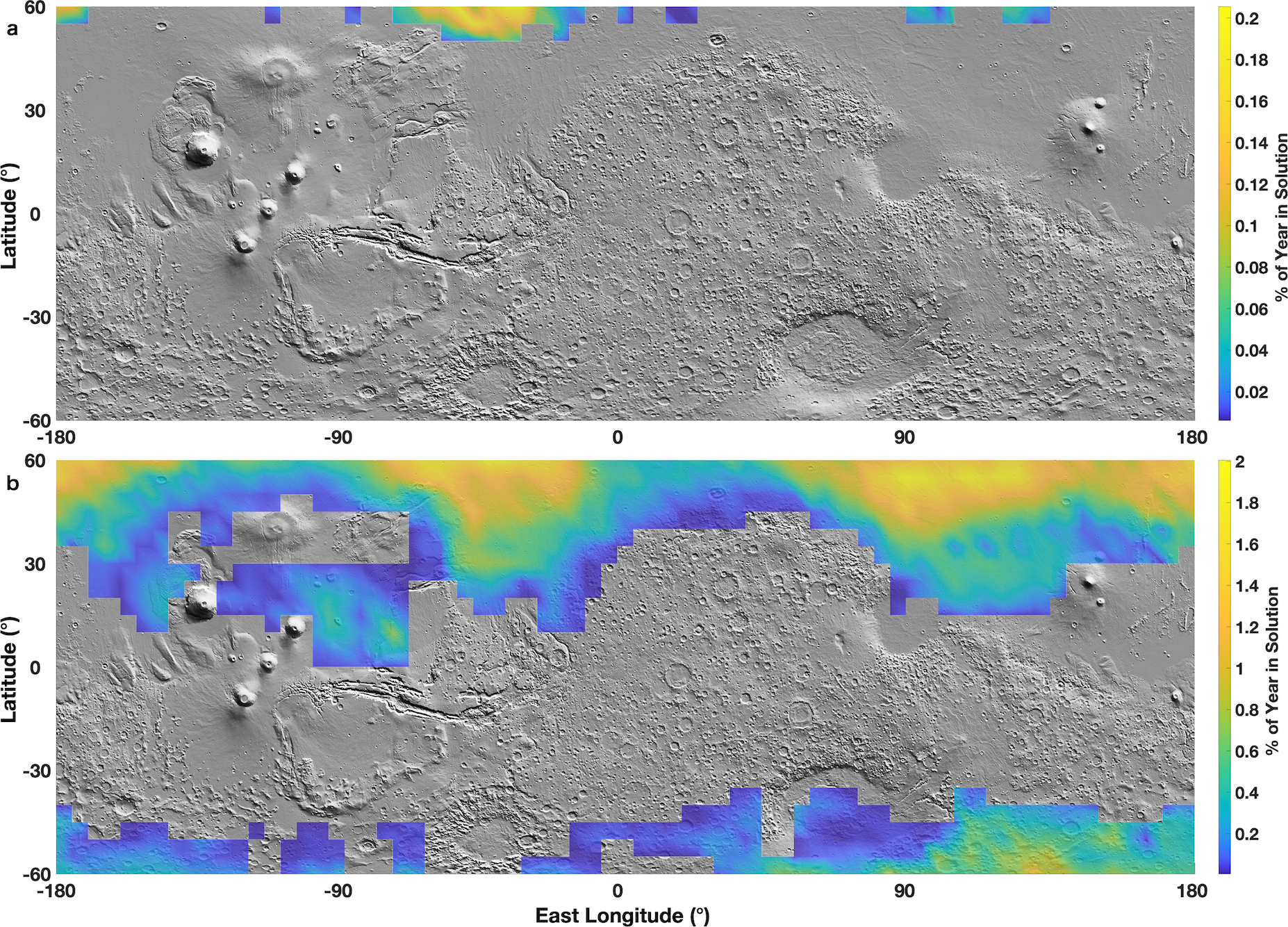}
\caption{\small{\textbf{Distribution of (meta)stable brines on the Martian surface.}  The total number of hours, in terms of percent of the Martian year (shown in color), a (a) magnesium perchlorate and (b) calcium perchlorate brine formed by deliquescence can exist on the Martian surface, as constrained by the MarsWRF predicted surface environment. Areas where such brines can not form and persist are shown as the background, gray, shaded relief map based on Mars Orbiter Laser Altimeter (MOLA) data. The latitude range is restricted to non-polar regions.}}
\end{figure}

\newpage

\begin{figure}
\includegraphics[width=\textwidth]{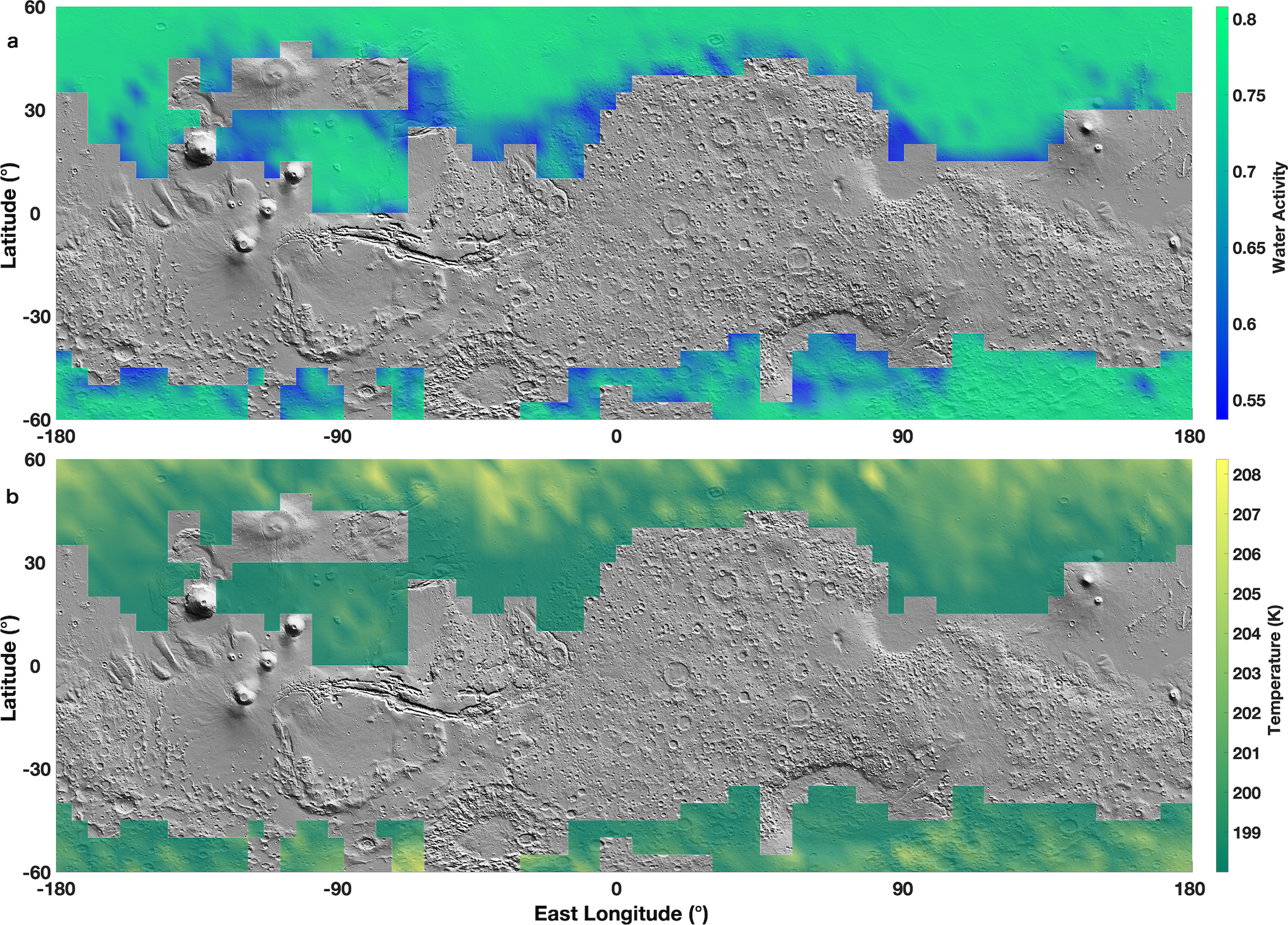}
\caption{\small{\textbf{Maximum achievable water activity and corresponding temperature of calcium perchlorate brines.} To better resolve the habitability of the most favorable liquid case, here the (a) maximum water activity achievable by brines formed through deliquescence of calcium perchlorate is shown along with the (b) corresponding brine temperature. The latitude range is restricted to non-polar regions. The background is the gray-scaled, shaded relief map of Mars based on MOLA data.} }
\end{figure}

\newpage

\begin{figure}
\includegraphics[width=\textwidth]{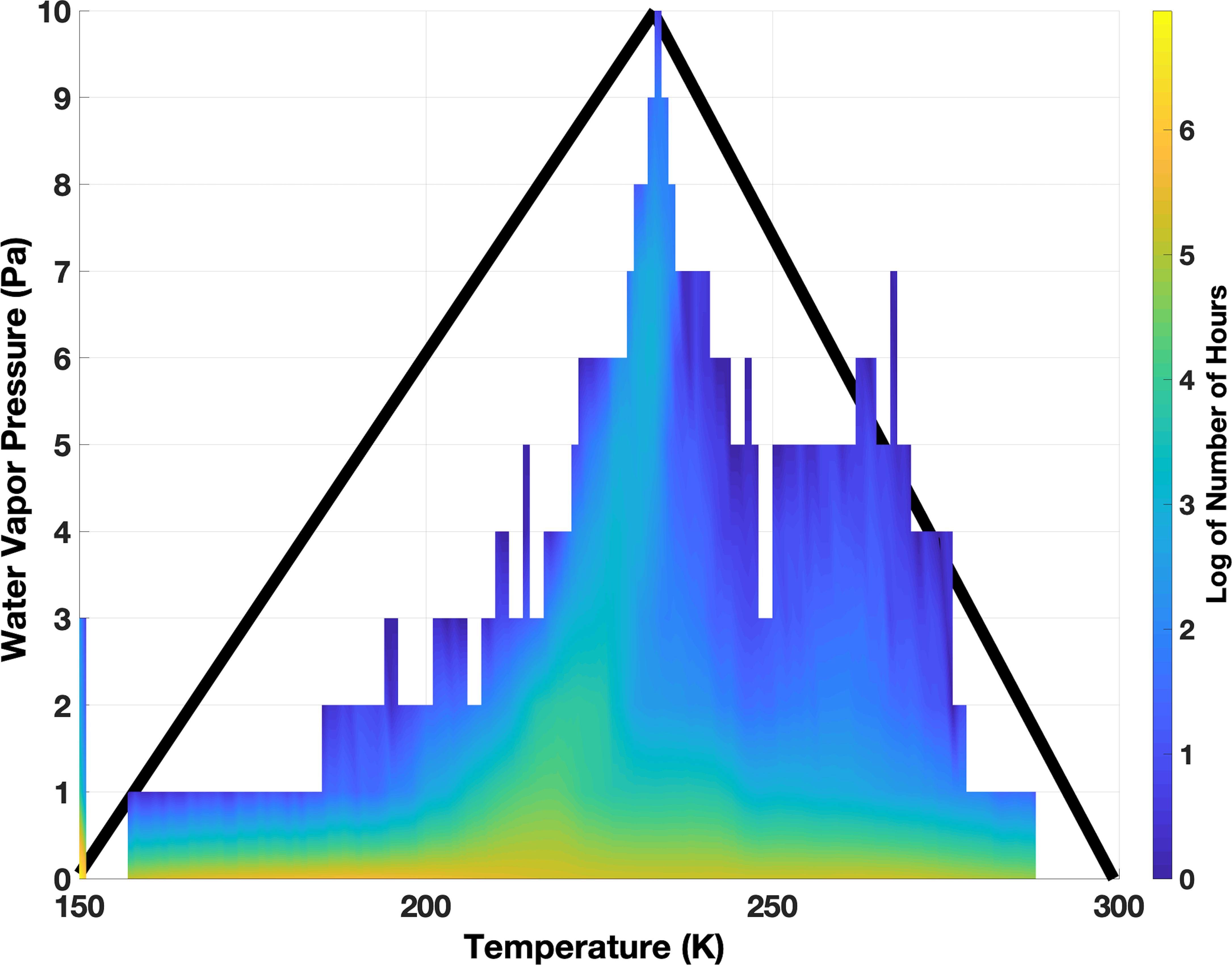}
\caption*{\small{\textbf{Extended Data Fig. 1 | Frequency of temperature and water vapor pressure pairings.} In Figure 1, the Mars-relevant pairings of surface temperature and atmospheric water vapor pressure are contained between the cyan lines. Here we show the frequency (in color) of each combination of surface temperature and water vapor pressure predicted by MarsWRF in support of the presented analysis in Figure 1. Frequency here is the log base 10 of the number of hours across the surface of Mars at a 5$^{\circ}$ $\times$ 5$^{\circ}$ latitude/longitude grid and one-hour temporal resolution. For the purposes of counting, temperature and water vapor pressure were binned in increments of 1~K and 1~Pa. Here the cyan lines from Figure 1 are in black.}}
\end{figure}

\newpage

\begin{figure}
\includegraphics[scale=0.17]{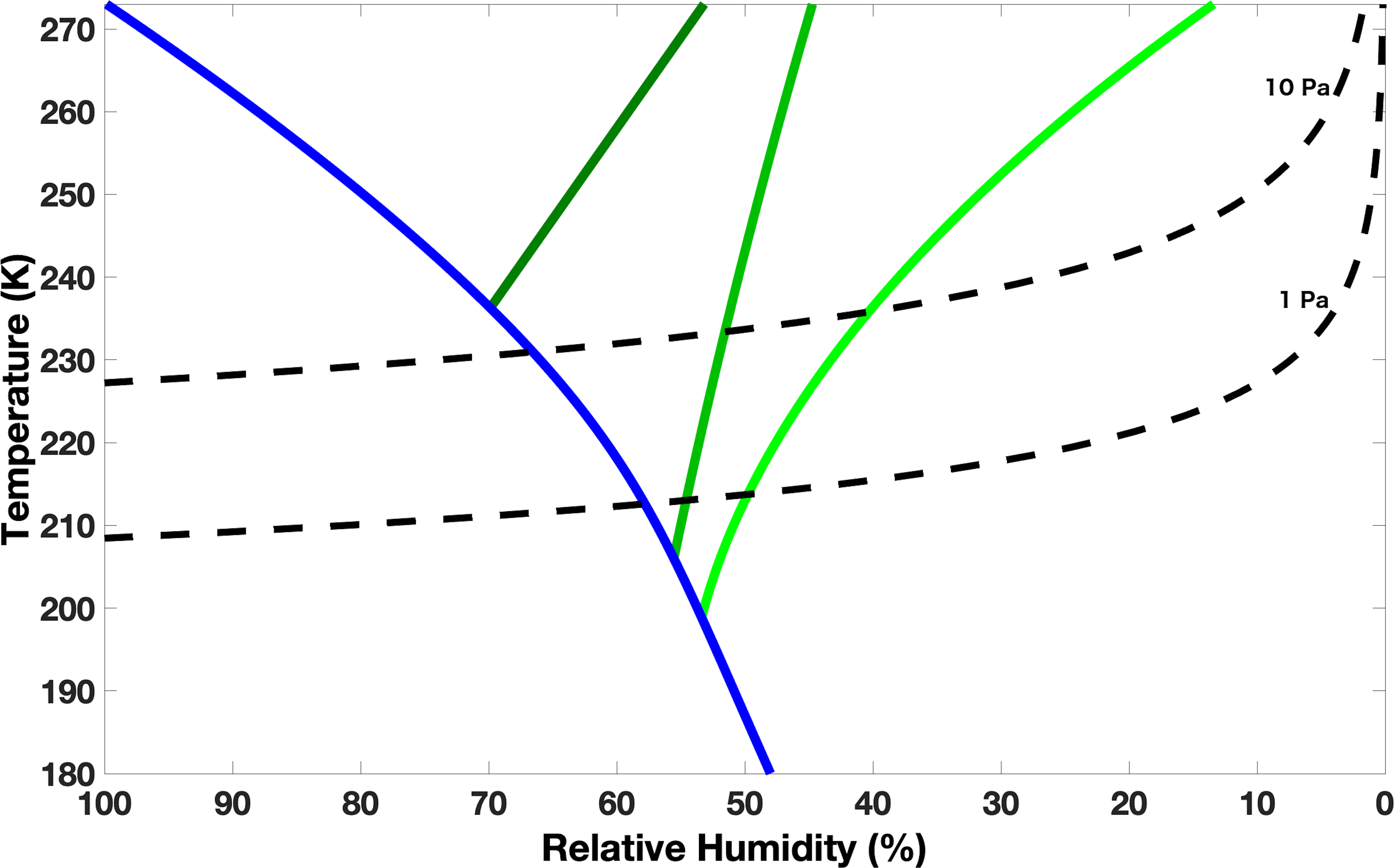}
\caption*{\small{\textbf{Extended Data Fig. 2 | Limits for Martian brine chemistries.} As per Figure 1, the maximum water activity of a brine that is thermodynamically stable on Mars is 0.66. This would imply that a brine with a eutectic water activity higher than this value would not readily form on present-day Mars. Another way of seeing this is in the phase diagram. Following the typical phase diagram (relative humidity vs temperature), here we show the ice line in blue (i.e., $RH_{ice} = 100$\%), the temperature-dependent deliquescence relative humidity (DRH) for calcium and magnesium perchlorate, as well as the sodium chlorate hydrate line in shades of green, from light to dark respectively. In dashed black lines, we plot two isobars for water vapor pressure, showing the typical maximum water vapor pressure measured on Mars by the Mars Science Laboratory rover and Phoenix lander, as well as the maximum surface water vapor pressure predicted by the MarsWRF model. The hyperarid conditions on Mars would not permit a salt with a eutectic water activity higher than 0.66 to form (i.e., eutectic temperature $> 230$~K). For example, at a eutectic temperature of 236~K, sodium chlorate would not form a brine because there is insufficient water vapor in the Martian atmosphere.}}
\end{figure}

\newpage

\begin{figure}
\includegraphics[scale=0.17]{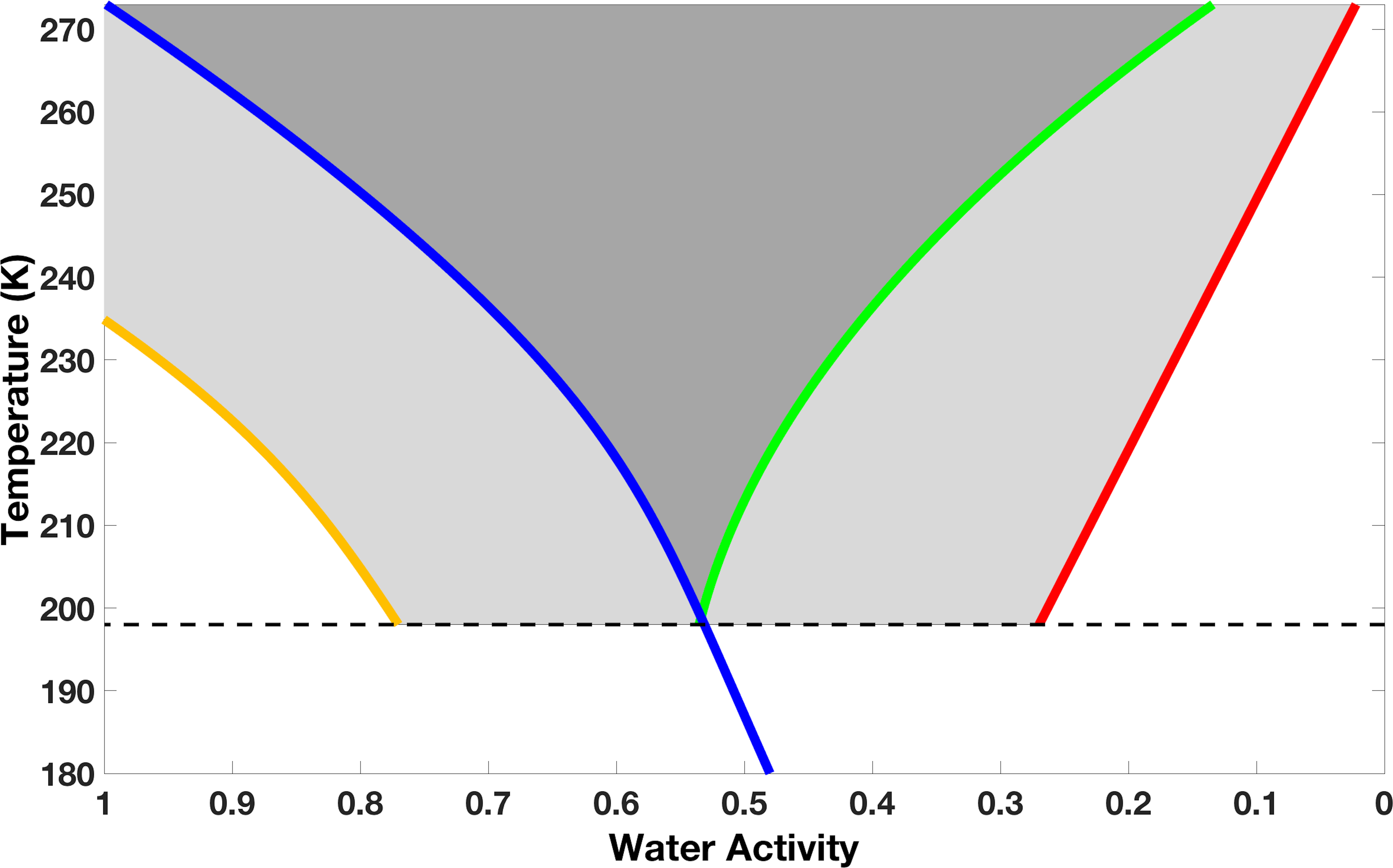}
\caption*{\small{\textbf{Extended Data Fig. 3 | Phase diagram of calcium perchlorate.} In Figures 2 and 3 we showed results for the (meta)stability of brines formed by the deliquescence of calcium and magnesium perchlorate. Here we show the thermodynamically stable (area shaded in dark gray) and metastable (area shaded in light gray) regions of a calcium perchlorate brine in the typical phase diagram. A brine can form between the ice line (i.e., $RH_{ice} = 100$\%), here the blue solid line, and the temperature-dependent deliquescence relative humidity (DRH), here shown as the green solid line. However, experiments have shown that metastable solutions exist beyond $RH_{ice} = 100$\%, up to $RH_{ice} = 145$\%, here shown as the orange solid line. Furthermore, although thermodynamically a solution should efflorescence once conditions fall below the DRH, experimental work has shown that solutions persist until much lower relative humidities are reached (i.e., the efflorescence relative humidity, here the red solid line). Non-shaded regions on this plot are conditions that would not permit for stable or metastable solutions of calcium perchlorate. The eutectic temperature ($\sim$198~K) is the black dashed line.}}
\end{figure}

\newpage

\begin{figure}
\includegraphics[width=\textwidth]{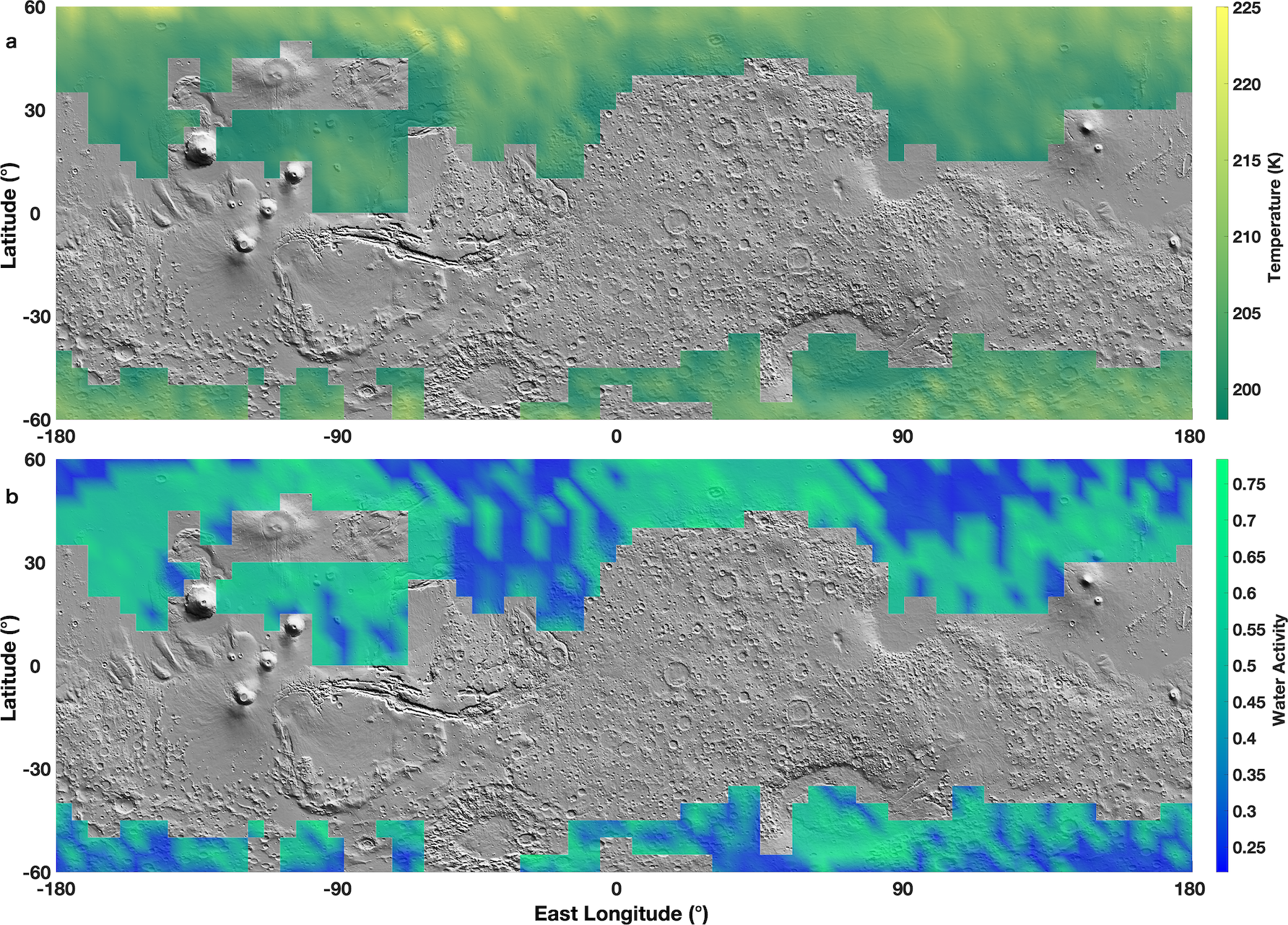}
\caption*{\small{\textbf{Extended Data Fig. 4 | Maximum achievable brine temperature and corresponding water activity of calcium perchlorate brines formed through deliquescence.} To better resolve the habitability of (meta)stable calcium perchlorate brines, here we plot (a) the maximum temperature and (b) corresponding water activity of resulting brines. The latitude range is restricted to non-polar regions. The background is the grey-scaled shaded relief map of Mars based on MOLA data.}}
\end{figure}

\newpage

\begin{figure}
\includegraphics[scale=0.12]{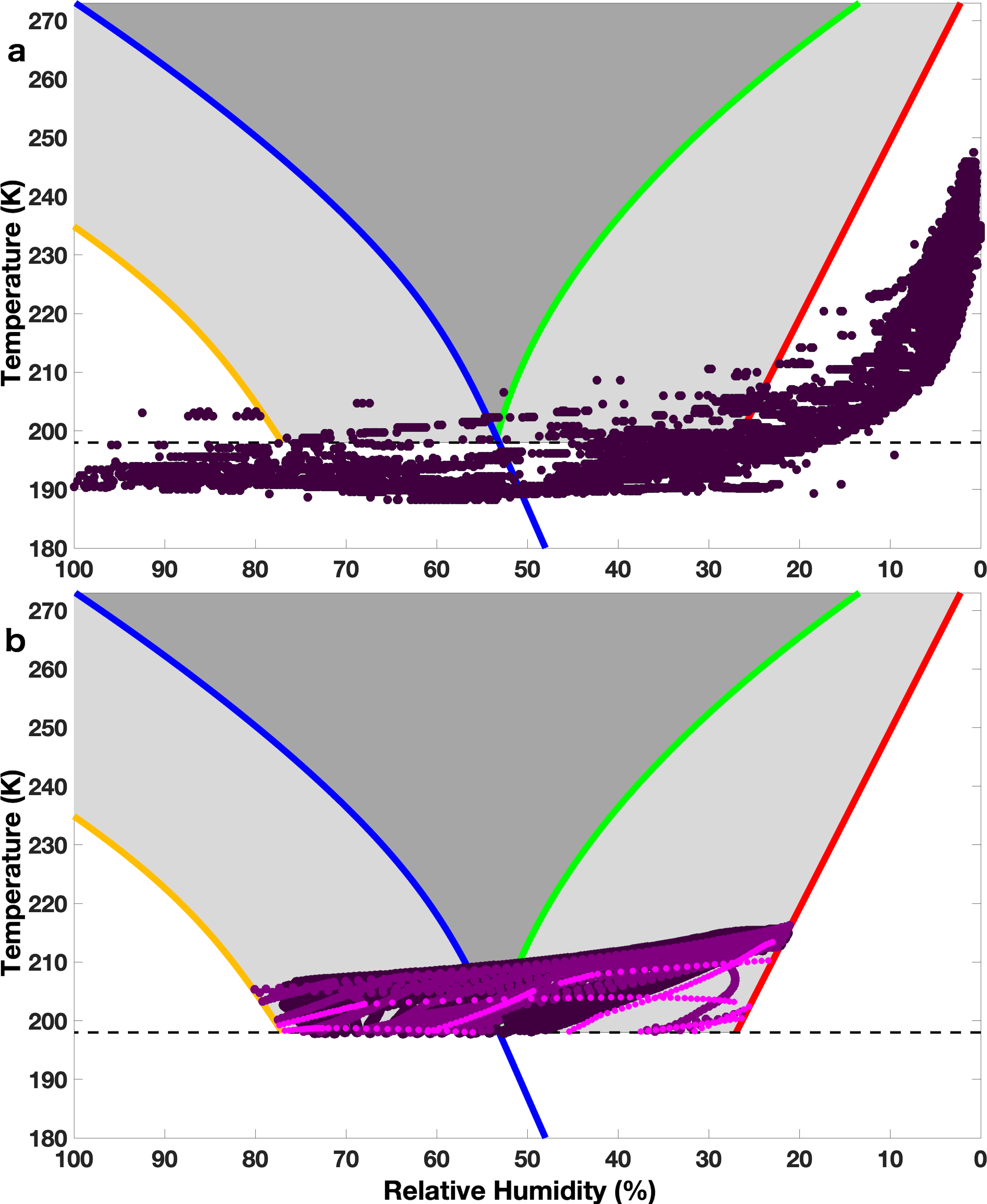}
\caption*{\small{\textbf{Extended Data Fig. 5 | Measured surface and modeled subsurface conditions at the Phoenix landing site.} Using newly recalibrated environmental data from the Thermal and Electrical Conductivity Probe (TECP) on the Phoenix Lander along with a model of the subsurface, we studied the potential to form liquids at the Phoenix landing site. In (a), we plot the Phoenix measured temperature and relative humidity with respect to liquid (purple points) on the phase diagram of calcium perchlorate (following the color code in Extended Data Fig.
3). As can be seen, several measured conditions are within the liquid stability region of calcium perchlorate brines. Furthermore, in (b) we plot the simulated subsurface conditions during which a brine is (meta)stable, assuming an ice table depth of 10 cm. The model’s surface predictions for both temperature and relative humidity were validated against the TECP measurements. In purple scale, from light to dark, we show results for 2, 5, and 8 cm.}}
\end{figure}

\end{document}